\documentclass[%
reprint,
showpacs,preprintnumbers,
nofootinbib,
 amsmath,amssymb,
 aps,
floatfix,
]{revtex4-1}
\usepackage[colorlinks=true,
urlcolor=cambridgeblue,
linkcolor=darkraspberry,
citecolor=cambridgeblue,
linktocpage=true,
pdfproducer=medialab,
pdfa=true,
anchorcolor=blue]{hyperref}%
\usepackage{xcolor}
\definecolor{cambridgeblue}{rgb}{0.64, 0.76, 0.68}
\definecolor{darkraspberry}{rgb}{0.53, 0.15, 0.34}
\usepackage{graphicx}
\usepackage{dcolumn}
\usepackage{bm}
\usepackage{hyperref}
\usepackage{amssymb}
\usepackage{placeins}
\usepackage[utf8]{inputenc}

\usepackage{float}
\usepackage[utf8]{inputenc} 
\usepackage{xcolor}
\usepackage{feynmp-auto}

\begin{document}

\title{Searches for light sterile neutrinos with multitrack displaced vertices}

\author{Giovanna Cottin}
\email{gcottin@phys.ntu.edu.tw}
\affiliation{Cavendish Laboratory, University of Cambridge, Cambridge CB3 0HE, UK}
\affiliation{Department of Physics, National Taiwan University, Taipei  10617, Taiwan}

\author{Juan Carlos Helo}
\email{jchelo@userena.cl }
  \affiliation{
Departamento de F\' isica, Facultad de Ciencias, Universidad de La Serena, 
Avenida Cisternas 1200, La Serena, Chile}
 \affiliation{ Centro-Cient\'\i fico-Tecnol\'{o}gico de Valpara\'\i so,
Casilla 110-V, Valpara\'\i so,  Chile.
}

\author{Martin Hirsch}
\email{mahirsch@ific.uv.es}
\affiliation{AHEP Group, Instituto de F\'{\i}sica Corpuscular --
    C.S.I.C./Universitat de Val{\`e}ncia, 
    Edificio de Institutos de Paterna, Apartado 22085,
    E--46071 Val{\`e}ncia, Spain}

\date{\today}


\begin{abstract}
We study discovery prospects for long-lived sterile neutrinos at the
LHC with multitrack displaced vertices, with masses below the
electroweak scale. We reinterpret current displaced vertex searches
making use of publicly available, parametrized selection efficiencies 
for modeling the detector response to displaced vertices. We focus on production 
of right-handed $W_{R}$ bosons and neutrinos $N$ in a left-right symmetric model, and find poor sensitivity.
After proposing a different trigger strategy (considering the prompt lepton 
accompanying the neutrino displaced vertex) and optimized cuts in the invariant 
mass and track multiplicity of the vertex, we find that the LHC with $\sqrt{s}=13$
TeV and 300 fb$^{-1}$ is able to probe sterile neutrino masses between
$10\hspace{0.1cm}\mbox{GeV}<m_{N}<20$ GeV (for a right-handed gauge
boson mass of $2\hspace{0.1cm}\mbox{TeV}<m_{W_{R}}<3.5$ TeV). To probe
higher masses up to $m_{N}\sim 30$ GeV and $m_{W_{R}}<5$ TeV, 3000
fb$^{-1}$ will be needed. This work joins other efforts in motivating
dedicated experimental searches to target this low sterile neutrino
mass region.

\end{abstract}

\maketitle

\section{Introduction}

With the discovery of the Higgs boson~\cite{Aad:2012tfa,Chatrchyan:2012xdj}, the Large Hadron Collider (LHC) has confirmed the particle content of the Standard Model (SM). Still one of the main unanswered fundamental question is the origin of neutrino masses. Experimental data on neutrino oscillations~\cite{Agashe:2014kda} provides clear evidence that there must be new physics beyond the SM, whose effects are actively being looked for by the LHC experiments.

The smallness of neutrino masses may be explained by the so-called see-saw mechanism~\cite{Minkowski:1977sc}, introducing the existence of massive, right-handed (i.e sterile) neutrinos~\cite{Mohapatra:1979ia,Schechter:1980gr} (see Ref.~\cite{Abazajian:2012ys} for a comprehensive review). Different realizations of this mechanism give rise to sterile neutrinos with Majorana masses covering various mass ranges (for a more extensive discussion on collider searches and limits, see Refs.~\cite{Cai:2017mow,Deppisch:2015qwa,Atre:2009rg} and references therein). Particularly in left-right symmetric extensions of the SM~\cite{Pati:1974yy,Mohapatra:1974gc}, production and decay of the sterile neutrino $N$ depends mostly on the unknown mass of the new, heavy right-handed gauge boson, $W_{R}$. 

In the mass region where $m_{N}\ll m_{W_{R}}$, the distinctive - lepton number violating - signature of same-sign dileptons~\cite{Keung:1983uu} has been extensively studied~\cite{Datta:1993nm,Panella:2001wq,Han:2006ip,delAguila:2007qnc,Basso:2008iv,Perez:2009mu,Atre:2009rg,Deppisch:2015qwa,Ng:2015hba,Cox:2017eme,Cai:2017mow,Das:2017gke}. If the sterile neutrino mass is also below the electroweak scale (i.e $m_{N}<m_{W}$), it can be long-lived, and travel a measurable distance before decaying inside the LHC detectors. 

Different collider searches for long-lived sterile neutrinos have been studied in Refs.~\cite{Dib:2018iyr,Antusch:2017hhu,Accomando:2017qcs,Dev:2017dui,Antusch:2016vyf,Antusch:2016ejd,Dube:2017jgo,Mitra:2016kov,Accomando:2016rpc,Izaguirre:2015pga,Gago:2015vma,Dib:2014iga,Helo:2013esa}. At the LHC, these include signatures of trileptons~\cite{Accomando:2017qcs,Izaguirre:2015pga}, lepton jets~\cite{Izaguirre:2015pga,Dube:2017jgo}, displaced ``neutrino jets"~\cite{Accomando:2017qcs,Mitra:2016kov}, displaced vertices from Higgs decays~\cite{Gago:2015vma,Accomando:2016rpc} and displaced vertices, defined in a broad sense~\cite{Helo:2013esa}. 

These distinct search proposals target different neutrino mass regions. For example, 13 TeV searches for 
``neutrino jets" can reach neutrino masses of order hundreds of GeV~\cite{Mitra:2016kov}. In this work we
are interested in $\mathcal{O}$ ($10$ GeV) neutrinos. Displaced lepton-jet searches can place 
strong limits on sterile neutrinos in the (4 GeV- 25 GeV) mass range~\cite{Dube:2017jgo}. Here we propose a 
complementary search for displaced vertices, that has the advantage of zero background.

This work considers a different search strategy, in which events are triggered by a prompt lepton and the neutrino charged decay products are identified to come from a common displaced vertex (DV), where a detailed detector response to displaced vertices is implemented in the form of parametrized selection efficiencies, recently made public by the ATLAS 
collaboration~\cite{SUSY-2016-08}. 

Our search proposal is inspired by the ATLAS multitrack displaced vertex analysis~\cite{Aaboud:2017iio,Aad:2015rba}. This work particularly expands the previous work in~\cite{Helo:2013esa} by considering a smaller fiducial region for displaced decay lengths, in order to have decays limited to the inner trackers of the LHC detectors ($<300$ mm) and by accurately modeling the detector response to displaced vertices inside them. In addition, our search strategy differs from the one in~\cite{Helo:2013esa} by triggering on the prompt lepton, instead of placing cuts on leptons and/or jets coming from the displaced vertex, and, by reconstructing the displaced vertex position from the neutrino charged decay products (i.e tracks). 

The rest of the paper is structured as follows. We briefly review the phenomenology of left-right symmetric models in Section~\ref{sec:model}. In Section~\ref{sec:SimResults} we discuss on our proposed DV strategy and discovery prospects at the LHC. Concluding remarks are presented in Section~\ref{sec:summary}. 

\section{Left-right symmetric model}
\label{sec:model}

The model considered is the left-right symmetric extension of the SM~\cite{Pati:1974yy,Mohapatra:1974gc,Mohapatra:1980yp}, with gauge group $SU(2)_{L}\times SU(2)_{R}\times U(1)_{B-L}$ and couplings $g_{L}, g_{R}, g_{1}$, respectively. This model contains a right-handed gauge boson $W_{R}$ and three right-handed Majorana neutrinos, with lightest state $N$.  In the mass range we are interested in (i.e  $m_{N}\ll m_{W_{R}}, m_{N}<m_{W} $), the sterile neutrino proper decay length can be written as~\cite{Helo:2013esa,Castillo-Felisola:2015bha,Maiezza:2015lza}

\begin{equation}
c\tau_{N}\sim 0.12\bigg (\frac{10 \hspace{0.1cm}\mbox{GeV}}{m_{N}}\bigg)^5\bigg(\frac{m_{W_{R}}}{1000 \hspace{0.1cm}\mbox{GeV}}\bigg)^4\hspace{0.2cm} [\mbox{mm}]
\label{eq:ctau}
\end{equation}

\begin{figure}[h]
\centering
\includegraphics{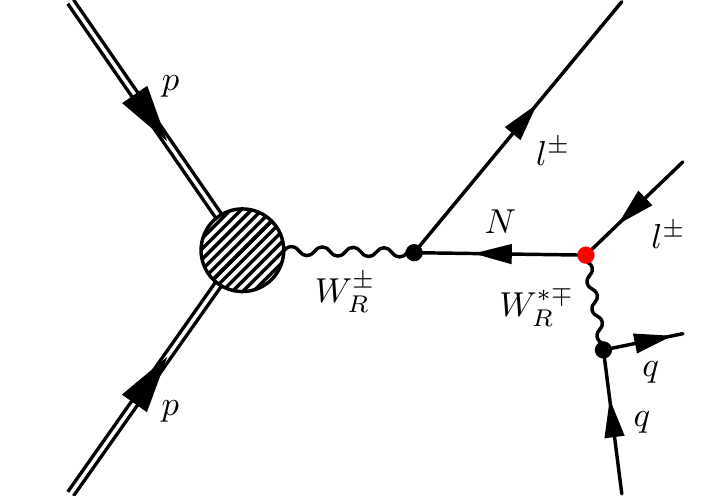}
\caption{Sterile neutrino $N$ production and decay at the LHC in the left-right symmetric model. For $m_{N}\ll m_{W_{R}}$, production proceeds through a $W_{R}$, followed by $N$ displaced decay to a lepton and quarks. The displaced vertex position is represented by the red circle. }
\label{FeynmanLRSM}
\end{figure}

A diagram showing sterile neutrino production and decay at the LHC can be seen in Figure~\ref{FeynmanLRSM}.

This process leads to different signatures, depending on 
the neutrino lifetime and mass. Existing searches by ATLAS~\cite{Aad:2015xaa} 
and CMS~\cite{Khachatryan:2014dka} place a lower limit on $m_{W_{R}}$ at 3
TeV, for $0.2\hspace{0.1cm}\mbox{TeV} \lesssim m_{N}\lesssim 2$
TeV. CMS has recently excluded right-handed bosons with masses
$m_{W_{R}}<4.4$ TeV, for $m_{N}=m_{W_{R}}/2$~\cite{CMS-PAS-EXO-17-011}.  
Searches for new resonances in the dijet distribution can also be used
to place limits on right-handed bosons 
\cite{ATLAS:2015nsi,Khachatryan:2015dcf,Aaboud:2017yvp}, roughly of
order 3 TeV.  For neutrino decays outside of the detector, searches for 
new heavy gauge bosons decaying to leptons and missing transverse 
momentum~\cite{Chatrchyan:2011dx,Aaboud:2017efa} can be sensitive.

Early constrains on left-right symmetric models from LHC data have been
addressed in~\cite{Nemevsek:2011hz}. The very recent study 
in~\cite{Nemevsek:2018bbt} tries to systematically asses constrains covering 
the entire neutrino mass range, and explores sensitivity from 
different searches. For decays in the ``displaced region" in the authors definition,
displaced jets searches (see for example~\cite{CMS:2014wda}) can be effective.

Currently, no public searches at the LHC target sterile neutrinos with 
masses as low as a few GeV. 

In the rest of the paper we restrict our discussion to sterile neutrino mixing 
with the electron sector only, for simplicity.

\section{Simulation and Results}
\label{sec:SimResults}

We generate a \textsc{UFO}~\cite{Degrande:2011ua} model with
\textsc{SARAH}~\cite{Staub:2013tta} and use
\textsc{SPheno}~\cite{Porod:2011nf,Porod:2003um} for the spectrum
calculation of the left-right symmetric model.
The \textsc{SARAH} model files are taken from the implementation
of the left-right symmetric model given in \cite{Bonilla:2016fqd}. 
We simulate events for
the process $pp\rightarrow W^{\pm}_{R}\rightarrow N
e^{\pm}$. Generation is performed with
\textsc{MadGraph5\_aMC@NLOv2.4.3}~\cite{Alwall:2014hca} at leading
order. The output corresponds to unweighted events in \textsc{LHE}
format~\cite{Alwall:2006yp}. The $N$ lifetimes are included 
using the \texttt{time\_of\_flight} option in \textsc{MadGraph5}.

The generated events are then interfaced to
\textsc{Pythia8 v2.15}~\cite{Sjostrand:2014zea} for hadronization and
computation of the $N$ decays. We use \textsc{FastJet
  3.1.3}~\cite{Cacciari:2011ma} for jet and trackless jet
reconstruction. Lepton reconstruction and the identification of the
displaced vertices is also done inside \textsc{Pythia8}. The masses
and decay widths of the particles in the model are communicated to
\textsc{Pythia} via the SLHA~\cite{Skands:2003cj,Allanach:2008qq}
section of the LHE header. Plots are generated with
{\texttt{matplotlib}}~\cite{Hunter:2007}.

The ATLAS analysis in Ref~\cite{Aaboud:2017iio} looks for high track-multiplicity displaced vertices at 13 TeV in events possessing displaced vertices and missing transverse momenta. We follow the detailed prescription using parametrized selection efficiencies as a function of displaced vertex radial distance, number of tracks and mass, that can be applied to vertices passing certain particle level acceptance requirements using the truth Monte Carlo event record. These efficiencies can be found in~\cite{SUSY-2016-08}, and are given for different regions in the detector, encapsulating also the effect of the material veto cut the analysis implements to remove background vertices~\footnote{The parametrized efficiencies provided by the ATLAS analysis were validated in Refs.~\cite{LLPwhitepaper,LLPLesHouches}.}. 

The analysis in Ref~\cite{Aaboud:2017iio} originally triggers on missing transverse momenta in the event to be bigger than $250$ GeV. In this work, we propose to trigger on the prompt lepton coming from the $W_{R}$ decay~\footnote{The ATLAS 8 TeV~\cite{Aad:2015rba} version of this analysis considers a DV+lepton signature, where the lepton that fires the trigger is associated with the displaced vertex. Here we do not follow this approach, as the decay products of the displaced $N$ are too collimated for the displaced lepton to satisfy isolation requirements. Alternatively, a ``trackless jet trigger"~\cite{Aad:2013txa} - where a jet matched with a muon is required - could be used to trigger on events, although it was shown that this trigger is inefficient for reconstructing
displaced vertices within the tracker~\cite{Csaki:2015fba}.}, as in our model there is little or no missing transverse momenta. Apart from the trigger requirement, we fully recast the multitrack analysis in~\cite{Aaboud:2017iio}. 

We require events to satisfy the following selections:

\begin{enumerate}

\item{Prompt electron: one electron\footnote{Note that in our simplified model we consider mixing in the electron sector only, so the prompt lepton is an electron. In the case where muon mixing is present, a muon trigger can be used.} with $p_{T}>25$ GeV.}

\item{Trackless jet: one ``trackless jet'' with $p_{T}>70$ GeV, or two trackless jets with $p_{T}>25$ GeV. A trackless jet is defined as a jet for which the scalar sum of the $p_{T}$ of all charged particles inside the jet does not exceed $5$ GeV~\footnote{This requirement is applied offline when processing data on disk, and is part 
of the first filtering in~\cite{Aaboud:2017iio} that gives the dataset where the {\it{large-radius
tracking}} is applied to. Therefore, this trackless jet cut can only see prompt tracks, so a particle 
decaying inside the tracker can pass this selection.}.}
\end{enumerate}

In addition, each event must have at least one displaced vertex with:

\begin{enumerate}
\setcounter{enumi}{2}
\item{DV fiducial: distance between the interaction point and the decay position $>4$ mm. The decay position must also lie in the fiducial region $r_{DV}<300$ mm and $|z_{DV}|<300$ mm.}

\item{DV $N_{trk}$: the number of selected decay products must be at least 5, where selected decay products are charged and stable, with $p_{T}>1$ GeV and transverse impact parameter $|d_{0}|>2$ mm.}

\item{DV $m_{DV}$: the invariant mass of the truth vertex must be larger than 10 GeV, and is constructed assuming all charged decay products have the mass of the pion.}

\item{DV efficiency: parametrized selection efficiencies are implemented depending on the displaced vertex distance, number of tracks and mass, following the prescription in~\cite{SUSY-2016-08}}.
\end{enumerate}

Based on these cuts, we first study the neutrino masses the analysis can be sensitive to. We simulate a grid of points with $m_{N}=[10 - 80]$ GeV and $m_{W_{R}}=[2 - 5]$ TeV. In each point we calculate the neutrino decay length, given by $c\tau_{N}\langle \beta \gamma \rangle$, where the average of the $\beta\gamma$ factor is taken after 10000 events, and it is roughly $\mathcal{O}(100)$ for the scanned grid. 

\begin{figure}
\includegraphics[width=0.5\textwidth,angle=0]{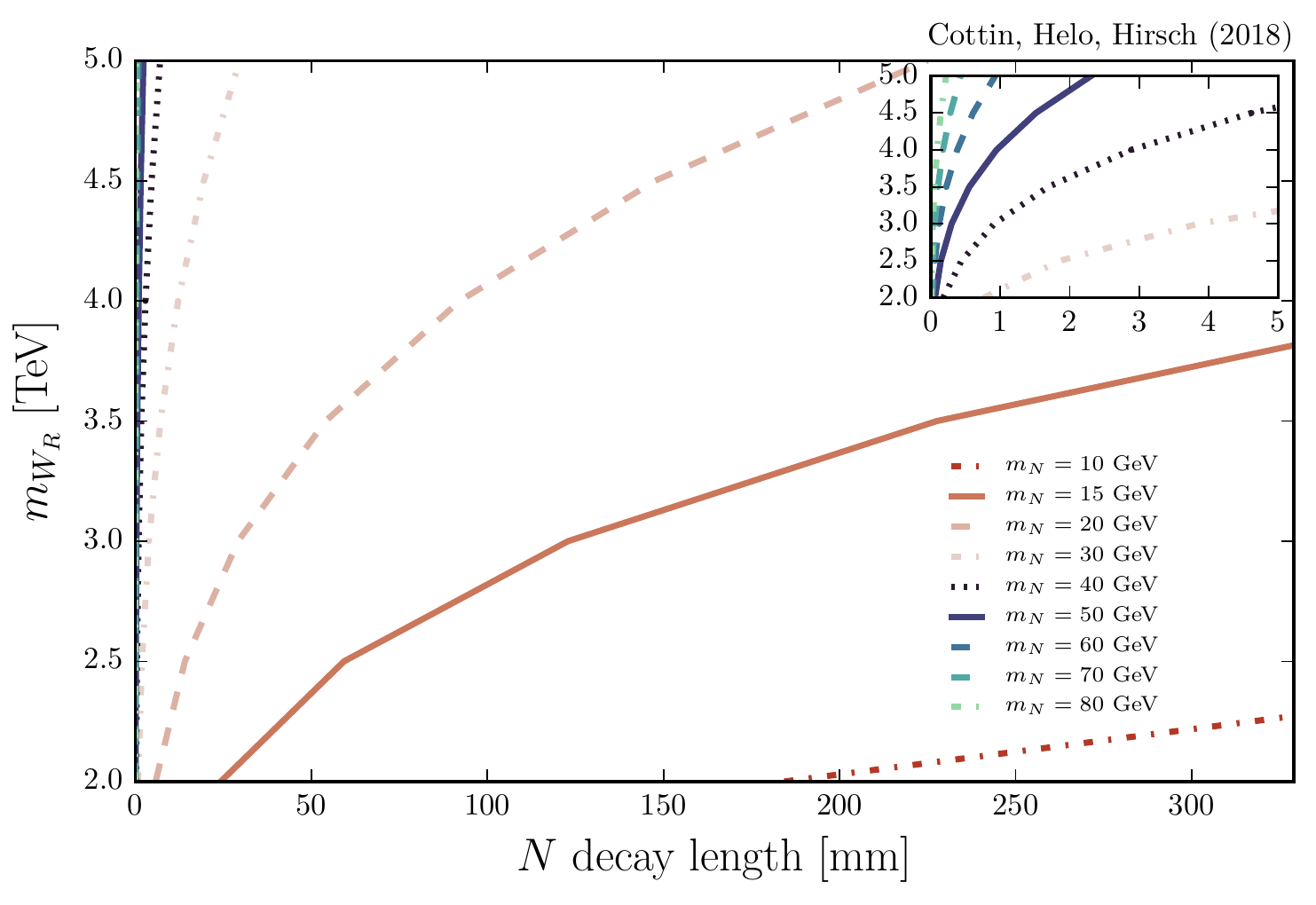}
\caption{Sterile neutrino decay distance as a function of $W_{R}$ mass, for fixed values of the sterile neutrino mass. The upper right plot shows a zoomed region, where higher neutrino mass curves can be seen.}
\label{Ndecay}
\end{figure}

Figure~\ref{Ndecay} shows the dependence on the decay position with the $W_{R}$ mass, for fixed values of the sterile neutrino mass. We note that for $m_{N}>40$ GeV, decays are below 4 mm, and will therefore fail to be in the fiducial region required by the analysis. An optimal acceptance region lies between $10\hspace{0.1cm}\mbox{GeV}<m_{N}<40$ GeV and $2\hspace{0.1cm}\mbox{TeV}<m_{W_{R}}<5$ TeV.

We therefore choose a particular benchmark point with $m_{N}=20$ GeV and $m_{W_{R}}=4$ TeV, and proper neutrino decay distance $c\tau_{N}=1.3$ mm. We show the effect of applying all analysis cuts on this benchmark in Table~\ref{tab:cutflow_Default}.

\begin{table}[h] 
\begin{tabular*}{0.45\textwidth}{@{\extracolsep{\fill}} l c c c}
\hline
                    & $N$          & Rel. $\epsilon$ [\%]   &   Ov. $\epsilon$ [\%] \\ 
\hline \hline 
All events          &    10000     &   100    &  100  \\ 
Prompt electron     &    8721      &   87.2   &  87.2  \\ 
Trackless jet       &    8704      &   99.8   &  87.0  \\ 
DV fiducial         &    7615      &   87.5   &  76.1  \\ 
DV $N_{trk}$        &    528       &   6.9    &  5.3  \\ 
DV $m_{DV}$         &    89        &   16.9   &  0.9  \\ 
DV efficiency       &    6         &   6.7    &  0.06  \\
\hline 
\end{tabular*} 
\caption{\label{tab:cutflow_Default}Numbers of simulated
  events $N$ at $\sqrt{s}=13$ TeV, relative and overall efficiencies for our left-right model with 
  $m_{N}=20$ GeV, $m_{W_{R}}=4$ TeV and $c\tau_{N}=1.3$ mm, for the ATLAS default cuts.}
\end{table} 

We see very low efficiency, reaching $0.06\%$ in Table~\ref{tab:cutflow_Default}. The loss comes mainly from the last two cuts: $N_{trk}\geq5$ and invariant vertex mass $m_{DV}>10$ GeV, which are too restrictive for the given neutrino mass of $m_{N}=20$ GeV, as softer decay products will lead to less amount of tracks available to make up a vertex. This sensitivity loss was also noted in a model with a long-lived singlino decaying to a light ($\sim20$ GeV) pseudoscalar~\cite{Allanach:2016pam}, and in a model with displaced Higgs decays to light ($\sim10$ GeV) hidden scalars~\cite{Csaki:2015fba}. 

We attempt to improve the sensitivity by loosening the $N_{trk}$ and $m_{DV}$ cuts, requiring $N_{trk}>3$ and $m_{DV}\geq5$ GeV. We should note that, even though the backgrounds to the multitrack displaced vertex search come mostly from instrumental sources (including hadronic interactions in dense material regions and random crossing of tracks), lowering the invariant mass and track multiplicity cuts could in principle rise background rates. Nevertheless, we expect these to still be within negligible levels with the requirements of displaced activity (i.e $d_{0}$ and trackless jet cut). In addition, there are still no background yields reported by ATLAS in this relaxed region, as it can be seen in Figure 7 of Reference~\cite{Aaboud:2017iio}. Of course, a dedicated experimental background estimation study with this optimized cuts would be needed, which is beyond the capabilities of our detector simulations. 

\begin{figure}
\includegraphics[width=0.55\textwidth,angle=0]{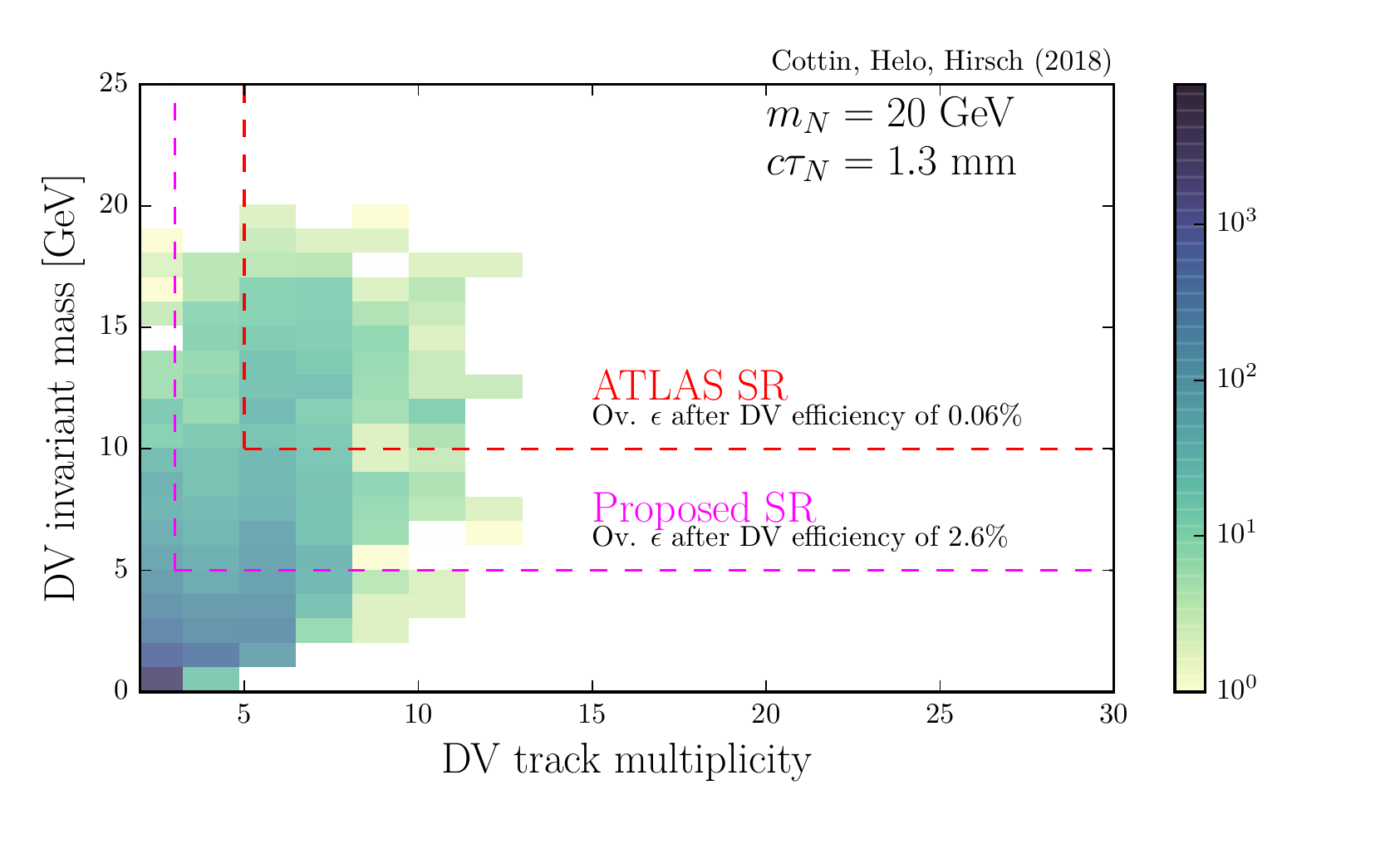}
\caption{Invariant mass of the displaced vertex against number of tracks for our left-right model with 
  $m_{N}=20$ GeV, $m_{W_{R}}=4$ TeV and $c\tau_{N}=1.3$ mm. Our proposed signal region is defined by the $N_{trk}>3$ and $m_{DV}\geq5$ GeV cuts.}
\label{mDVNtrk}
\end{figure}

\begin{figure*}
\includegraphics[width=0.8\textwidth,angle=0]{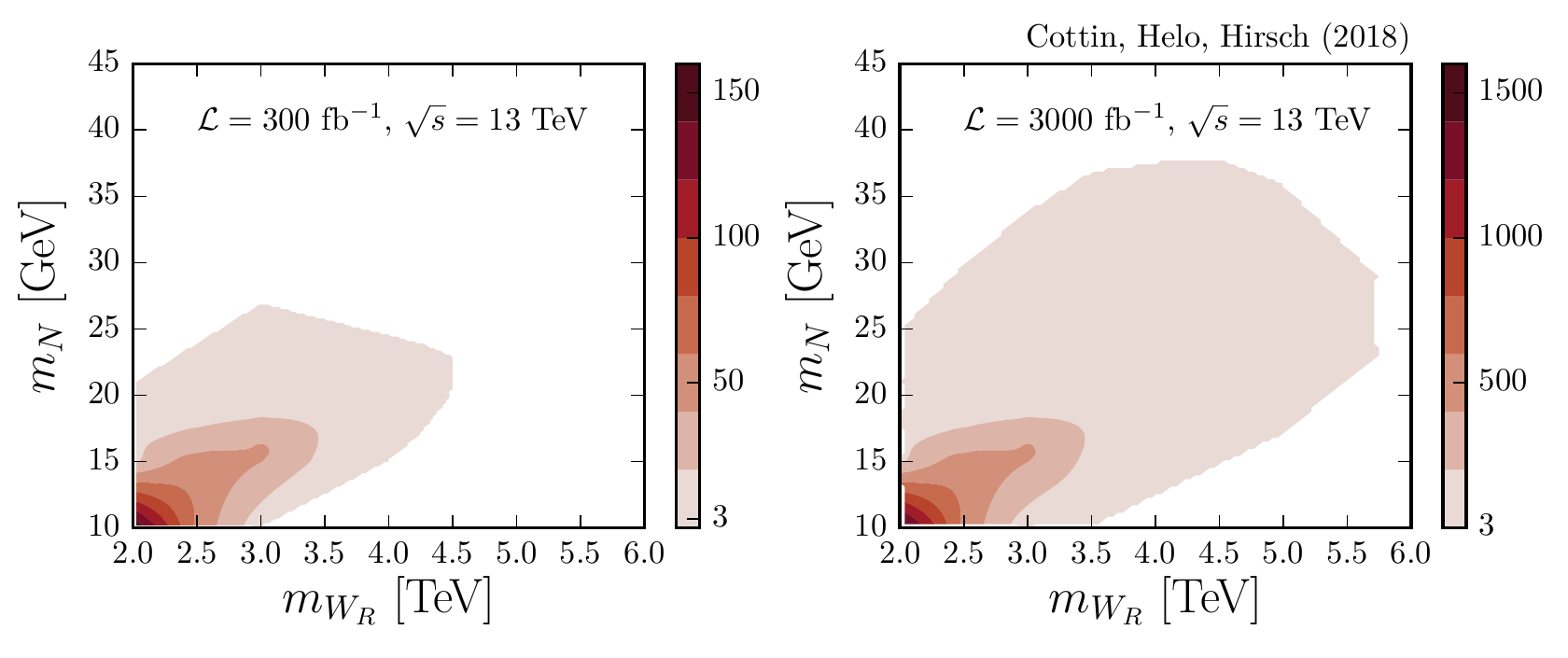}
\caption{Number of signal events for our left-right model at $\sqrt{s}=13$ TeV, with our proposed search strategy, with $\mathcal{L}=300$ fb$-1$ (left) and $\mathcal{L}=3000$ fb$-1$ (right).}
\label{NsignalLR}
\end{figure*}

Figure~\ref{mDVNtrk} shows the invariant mass of the displaced vertex against number of tracks for our benchmark. 
Prompt electron, Trackless jet and DV fiducial cuts are applied (cuts 1. to 3.). The last two cuts of DV $N_{trk}$ and DV $m_{DV}$ (cuts 4. and 5.) define the regions where signal is expected to be found (SR) and are shown as dashed boxes in the Figure. The red box defines a SR with the default ATLAS cuts of $N_{trk}\geq 5$ and $m_{DV}>10$ GeV. The purple box defines our proposed SR with $N_{trk}>3$ and $m_{DV}\geq5$ GeV. On each boxed region we show the final efficiency obtained after applying the experimental DV efficiency (cut 6.). We see an efficiency improvement of $2$ orders of magnitude in our proposed SR.

With the proposed tuned cuts of $N_{trk}>3$ and $m_{DV}\geq5$ GeV we now discuss discovery prospects. Figure~\ref{NsignalLR} shows the number of signal events in the mass plane $m_{W_{R}}-m_{N}$. Regions producing at least 3 signal events are shown, which is reasonable to set as
a requirement for discovery in the absence of background. We see in the left plot of Figure~\ref{NsignalLR} that at 13 TeV and with 300 fb$^{-1}$ the LHC is sensitive to  masses up to $m_{N}\sim 20$ GeV, for $m_{W_{R}}\sim 3.5$ TeV. With a higher luminosity of 3000 fb$^{-1}$ in the right plot, the 13 TeV LHC can reach sterile neutrino masses up to $\sim 30$ GeV for similar $W_{R}$ masses.

\section{Summary and Conclusions}
\label{sec:summary}

We have studied the sensitivity of current multitrack displaced vertex searches at the LHC for probing long-lived, light sterile neutrinos. We based our study on a left-right symmetric model, and consider sterile neutrino masses $m_{N} \ll m_{W_{R}}$, and $m_{N}<m_{W}$. 

We find that for $40\hspace{0.1cm}\mbox{GeV}<m_{N}<80$ GeV, neutrino decay distances are below 4 mm, and will therefore fail to be in the analysis fiducial tracker region. 

After considering a different trigger strategy on the prompt lepton in the event, we find very poor signal efficiency for events passing the standard DV cuts. This is due to the low mass of the sterile neutrino, being too soft to produce much tracks. 

After loosening the DV invariant mass and number of tracks to a region where still zero background is reported~\cite{Aaboud:2017iio}, we find that with $3000$ fb$^{-1}$ of integrated luminosity at 13 TeV, this proposed ``loose multitrack DV + prompt lepton" search should be sensitive to sterile neutrino masses up to $\sim 30$ GeV. 

The identification of displaced vertices will continue to be an important signature of new physics, given that this signal is scarce in the SM. We encourage the experimental collaborations at the LHC to pursue this searches further, particularly in the study of lower cuts in the displaced vertex invariant mass, as $m_{N}$ lower than 10 GeV may be probed.

\acknowledgments{G.C. thanks Christian Ohm for helpful discussions on
  the ATLAS multitrack displaced vertex analysis. The work of G.C.
  while at Cambridge was funded by the postgraduate
  Conicyt-Chile/Cambridge Trusts Scholarship 84130011, where this work
  was initiated. G.C. is now supported by the Ministry of Science and
  Technology of Taiwan under grant No. MOST-106-2811-M-002-035. J.C.H.
  is supported by Chile grants Fondecyt No. 1161463, Conicyt PIA/ACT
  1406 and Basal FB0821. M. H. was funded by Spanish MICINN grant
  FPA2017-85216-P and SEV-2014-0398 (from the Ministerio de Economía,
  Industria y Competitividad), as well as PROMETEOII/2014/084 (from
  the Generalitat Valenciana).}

\FloatBarrier

\bibliographystyle{apsrev4-1}
\bibliography{main}

\end{document}